# Frequency conversion efficiency of CO- and $CO_2$-laser radiation in $BaGa_2GeSe_6$ crystal

**IONIN A.A.[1], KINYAEVSKIY I.O.[1,*], MOZHAEVA V.A.[1,2]**

[1]*P.N. Lebedev Physical Institute of the Russian Academy of Sciences, Moscow, 119991, Russia*
[2]*Moscow State University of Geodesy and Cartography, Moscow, 105064, Russia*
[*]*kigor@sci.lebedev.ru*

**Abstract:** Non-linear optical characteristics of a new $BaGa_2GeSe_6$ crystal were numerically studied and compared with well-known mid-IR nonlinear crystals $ZnGeP_2$, GaSe, and $AgGaSe_2$. The calculations demonstrated for the new crystal to be probably the most efficient crystal for second harmonic generation of CO- and $CO_2$-laser radiation. It was found that a broadband two-stage frequency conversion of multi-line CO-laser radiation in this crystal is possible within 2.5-9.0 μm wavelength range with higher efficiency than in $ZnGeP_2$ and $AgGaSe_2$ crystals.

## 1. Introduction

A number of fundamental and applied problems in the fields of photochemistry, spectroscopy, plasma diagnostics and others need mid-IR laser sources. To obtain mid-IR laser radiation, different nonlinear techniques have been extensively used: optical parametric oscillators and difference frequency generation of short-wave (0.7–3 μm) lasers [1]; sum and difference frequency generation of mid-IR lasers [2]. Nonlinear optical crystals should have following properties: transparency in required spectral range; adequate birefringence for phase-matching; high nonlinear coefficient; high optical damage threshold and some others. The most known and widespread mid-IR crystals are $ZnGeP_2$, GaSe, and $AgGaSe_2$ [3, 4]. It is worth noticing that the $ZnGeP_2$ crystal is usually considered to be the most efficient one and sometimes called the "standard" of mid-IR crystals [4].

A new promising uniaxial nonlinear crystal $BaGa_2GeSe_6$ (BGGSe) was developed in the Kuban State University [5]. The BGGSe crystal has a high optical damage threshold (110 MW/cm$^2$), a wide spectral transparency range (from 0.5 to 18 μm) and a high nonlinearity ($d_{11} = 66 \pm 15$ pm/V) [5], which makes this crystal very attractive for frequency conversion of laser radiation into (or within) the mid-IR range. Therefore, the objective of this paper is estimation of BGGSe crystal efficiency with respect to $ZnGeP_2$, GaSe, and $AgGaSe_2$ by means of frequency conversion simulation for CO- and $CO_2$-laser radiation. The calculations under discussion demonstrated that broadband two-stage frequency conversion of multi-line CO-laser is quite possible in a BGGSe crystal as well as in $ZnGeP_2$ [6] and $AgGaSe_2$ [7].

## 2. Second harmonic generation efficiency

The dependence of phase-matching angle $θ$ for I-type second harmonic generation (SHG) in a BGGSe crystal versus pump radiation wavelength is presented in Fig. 1. Spectral bands of a CO-laser (from 4.7 to 8.7 μm) and $CO_2$-laser (from 9.2 to 10.8 μm) are also indicated in Fig. 1. The calculation was performed by the dispersion equation taken from [5].

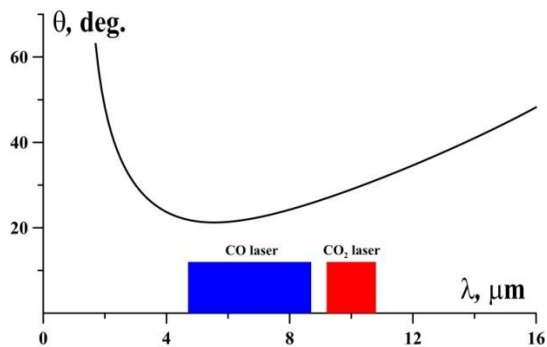

Fig. 1. The dependence of the phase-matching angle of second harmonic generation in a BGGSe crystal versus pump radiation wavelength.

Fig. 1 shows that phase-matched SHG is possible in the wide interval of mid-IR range including spectral bands of CO- and $CO_2$-lasers. The minimum of the dependence in Fig. 1 is located at 5.5 μm wavelength that belongs to the spectral band of a CO-laser. This means that noncritical spectral phase-matching for CO-laser radiation takes place, and broadband frequency conversion of multi-line CO-laser radiation is possible likewise in $ZnGeP_2$ [6] and $AgGaSe_2$ crystals [7].

To compare the efficiency of BGGSe crystal with common $ZnGeP_2$, GaSe, and $AgGaSe_2$ crystals, the SHG efficiency ($K$) was numerically calculated for 5-μm (CO-laser) and 10.6-μm ($CO_2$-laser) radiation. The calculation was carried out in the approximation of plane waves and exact phase-matching under the nonlinear conversion regime by formula [3]:

$$K = th^2\left(\frac{L}{L_{nl}}\right), \quad (1)$$

where $L$ – the crystal length, $L_{nl}$ – the length of nonlinear interaction (the length corresponding to the conversion efficiency of ≈58%), calculated using formula [3]:

$$L_{nl} = \frac{n_3 \lambda_1}{2\pi^2 d_{eff}} \cdot \left(\frac{n_1}{752 \cdot I}\right)^{\frac{1}{2}}, \quad (1)$$

where $n_3$, $n_1$ – the refractive indices for SHG and pump radiation, respectively, $\lambda_1$ – the wavelength of pump radiation, $d_{eff}$ – the effective nonlinear coefficient and $I$ – the intensity of pump radiation. A difference of optical strength of the crystals was taken into account by carrying out the calculation assuming intensity equaled to the corresponding optical damage threshold. The optical damage threshold depends on the laser emission wavelength and pulse duration; therefore we selected measured data for similar conditions: the radiation wavelength was ~ 10 μm, the pulse duration was ~ 0.1 μs. The properties of $ZnGeP_2$, GaSe, and $AgGaSe_2$ crystals were taken from reference [3], and of BGGSe crystal from [5]. The input and output data of the calculation are presented in Table 1.

**Table 1. The input and the output data of the calculation**

| Crystal | d, pm/V | I, MW/cm² | $\theta_{CO}$, deg | $L_{nl}^{CO}$, mm | $\Theta_{CO2}$, deg | $L_{nl}^{CO2}$, mm |
|---|---|---|---|---|---|---|
| $AgGaSe_2$ | $d_{36}$=33 | 20 | 42.5 | 3.92 | 55.5 | 6.78 |
| GaSe | $d_{22}$=54 | 30 | 10.8 | 1.42 | 14.9 | 3.01 |
| $BaGa_2GeSe_6$ | $d_{11}$=66 | 110 | 21.6 | 0.60 | 30.6 | 1.47 |
| $ZnGeP_2$ | $d_{36}$=75 | 78 | 48.0 | 0.77 | 76.0 | 3.37 |

The calculated dependences of SHG efficiency on the crystal length are presented in Fig. 2 for CO-laser (a) and $CO_2$-laser (b) emission. Fig. 2 shows that the BGGSe crystal is the most efficient one among the studied crystals under the considered conditions, because the conversion of pump radiation into SHG radiation occurs along the shortest length. The calculation shows that the BGGSe crystal is even more efficient than the $ZnGeP_2$ crystal. The least efficient among considered crystals was the $AgGaSe_2$.

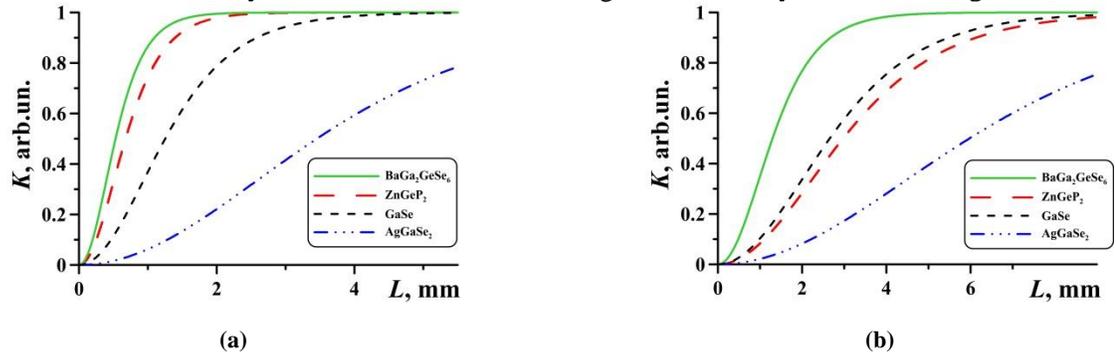

(a)  (b)

Fig. 2. The dependences of SHG efficiency on the length of nonlinear crystals for CO-laser (a) and CO2-laser (b) radiation.

It should be noted that some factors affecting the conversion efficiency were not considered in the calculation, such as absorption, intensity losses due to Fresnel reflection from the crystal facets and a walk-off effect. The walk-off effect can significantly decrease the conversion efficiency in a crystal with high birefringence as a GaSe crystal (see, for instance, [8]). However, the walk-off effect can be neglected for the wide-aperture beams.

### 3. Broadband two-stage frequency conversion of the multi-line CO laser emission

As follows from Fig. 1, that noncritical spectral phase-matching takes place for SHG of CO-laser in BGGSe crystal. This fact indicates the possibility of a broadband two-stage frequency conversion of multi-line CO laser radiation, which was previously performed in $ZnGeP_2$ [6] and $AgGaSe_2$ crystals [7]. The highest conversion efficiency and the widest spectral band of the two-stage frequency conversion were obtained in a $ZnGeP_2$ crystal. Therefore, we carried out a numerical simulation of the broadband two-stage frequency conversion of the multi-line CO-laser in BGGSe and ZnGeP2 crystals under the same conditions.

The first stage of the frequency conversion is a broadband sum frequency generation (SFG) of multi-line CO laser radiation; the second one is a broadband difference frequency generation (DFG) between SFG radiation (the first cascade radiation) and the rest of the pump radiation. SFG and DFG processes occur concurrently in the same crystal in the same direction wherein SFG and DFG spectrum consists of ~$10^4$ and $10^6$ lines, respectively. Therefore, the simulation of broadband two-stage frequency conversion is very complicated problem which can be significantly simplified in the case of low-efficiency approximation. In this case, the power of each spectral line can be calculated independently of others, that allows one to avoid the problem of a big system of differential equations. The procedure of broadband two-stage frequency conversion of the multi-line CO-laser radiation under the low-efficiency approximation is described in details in [6].

The simulation was carried out for multi-line CO-laser radiation described in [6]: pulse duration ~1 µs, peak power (integrated over spectrum) 4 kW, the radiation spectrum contained ~150 lines in the wavelength interval from 5 to 7.5 µm (Fig. 3). The radius of laser beam was 0.5 mm, the effective length was 5 mm for both $ZnGeP_2$ and BGGSe crystals. The phase-matching angles of $ZnGeP_2$ and BGGSe crystals were 48° and 22°, respectively, corresponding to the phase-matching angles for SHG with the wavelength of 2.5 µm.

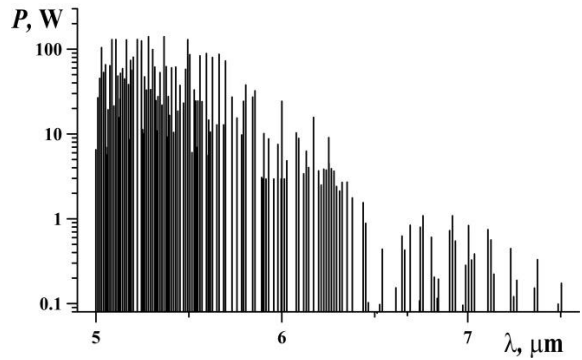

Fig. 3. The spectrum of CO laser taken from [6].

The calculated spectra of SFG radiation (the first stage) are presented in Fig. 4. The spectra are clipped at 1 mW peak power level. The conversion efficiency integrated over the spectrum was 3.6% and 0.26% for BGGSe and $ZnGeP_2$ crystals, respectively, which corresponded to low-efficiency approximation. Fig. 4 shows that the SFG spectrum obtained in BGGSe covers a wide wavelength interval from 2.5 to 3.2 µm spreading wider than that of $ZnGeP_2$ crystal. The peak power of the SFG radiation in BGGSe crystal is also higher than that of $ZnGeP_2$ crystal. The peak power of the strongest SFG lines in BGGSe and $ZnGeP_2$ crystals was 0.48 W and 0.27 W, respectively. Thus, we can conclude that under considered conditions the new crystal is approximately twice more efficient than the $ZnGeP_2$ crystal.

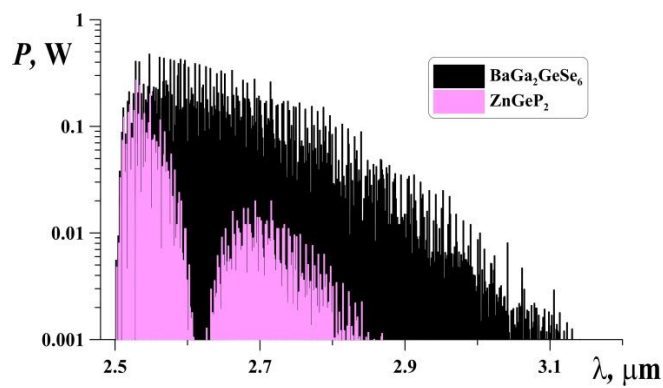

Fig. 4. The spectra of SFG radiation calculated for ZnGeP2 and BGGSe crystals.

In the second stage of the frequency conversion, SFG radiation interacted with the pump radiation resulting in DFG. The phase-matching angle was the same. Calculated spectra of DFG for $ZnGeP_2$ and BGGSe crystals are presented in Fig. 5. The spectra are clipped at 1 nW peak power level.

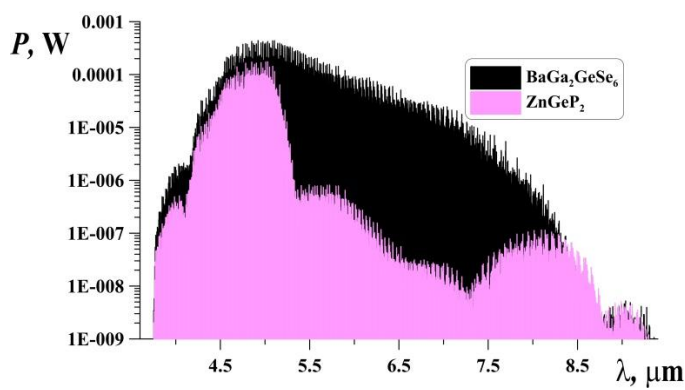

Fig. 5. The spectra of DFG radiation calculated for ZnGeP2 and BGGSe crystals.

The DFG spectra for both crystals cover the spectral interval from 4 to 9 μm quite densely. The peak power of DFG radiation in BGGSe crystal (up to 0.35 mW) is higher than that of $ZnGeP_2$ crystal (up to 0.16 mW) in the whole considered spectral range. Hence, we can come to the conclusion that under considered conditions the conversion efficiency of second stage in BGGSe crystal is at least 2 times higher than that of $ZnGeP_2$ crystal. Thus, for the broadband two-stage frequency conversion of multi-line CO laser radiation, the new BGGSe crystal is twice more efficient than "standard" nonlinear $ZnGeP_2$ crystal.

## 4. Summary

A numerical study of the new $BaGa_2GeSe_6$ crystal was performed in comparison with well-known mid-IR nonlinear crystals such as $ZnGeP_2$, GaSe, and $AgGaSe_2$. The calculations demonstrated that the $BaGa_2GeSe_6$ crystal proved to be more effective, even than $ZnGeP_2$ crystal for second harmonic generation of CO-laser and $CO_2$-laser radiation. It looks like this crystal is very attractive for mid-IR optical parametric oscillators and phase-matched generation/amplification of ultra-short mid-IR pulses as in reference [9]. The noncritical spectral phase-matching takes place for fundamental vibrational band CO-laser, and broadband two-stage frequency conversion of multi-line CO-laser radiation is possible within 2.5-9.0 μm wavelength range. The efficiency of broadband two-stage frequency conversion in $BaGa_2GeSe_6$ crystal is at least twice higher than in $ZnGeP_2$ crystal.

## 5. Funding

The study was supported by the Russian Science Foundation grant №16-19-10619.